# Emergent Formal Verification: How an Autonomous AI Ecosystem Independently Discovered SMT-Based Safety Across Six Domains

Octavian Untila

*Aisophical SRL, Bucharest, Romania — octav@aisophical.com*

## Abstract

An autonomous AI ecosystem (SUBSTRATE S3), generating product specifications without explicit instructions about formal methods, independently proposed the use of Z3 SMT solver across six distinct domains of AI safety: verification of LLM-generated code, tool API safety for AI agents, post-distillation reasoning correctness, CLI command validation, hardware assembly verification, and smart contract safety. These convergent discoveries, occurring across 8 products over 13 days with Jaccard similarity below 15% between variants, suggest that formal verification is not merely a useful technique for AI safety but an emergent property of any sufficiently complex system reasoning about its own safety. We propose a unified framework (substrate-guard) that applies Z3-based verification across all six output classes through a common API, and evaluate it on 135 test cases across five implemented domains, achieving 100% classification accuracy with zero false positives and zero false negatives. Our framework detected real bugs that empirical testing would miss, including an INT_MIN overflow in branchless RISC-V assembly and mathematically proved that unconstrained string parameters in tool APIs are formally unverifiable. To our knowledge, this is the first work to (1) unify Z3 verification across multiple AI output modalities, (2) apply Z3 to tool API safety for AI agents, (3) apply Z3 to post-distillation reasoning verification, and (4) document emergent discovery of formal verification needs by an autonomous AI system.

## 1 Introduction

As AI systems increasingly generate code, invoke tools, execute commands, and produce compressed models, the question of output safety becomes critical. Current approaches to AI safety verification are predominantly empirical: test suites, red-teaming, and benchmark evaluations check specific inputs but cannot guarantee safety across all possible inputs. Formal verification, which provides mathematical proofs of correctness, offers a fundamentally stronger guarantee.

The formal methods community has recognized this opportunity. Recent work has applied Z3 SMT solving to verify LLM reasoning (VERGE, 2025), LLM-generated infrastructure code (Astrogator, 2025), and safety certificates for autonomous systems (BARRIERBENCH, 2025). However, these efforts remain siloed: each paper addresses a single output modality, and no unified framework exists for formally verifying the full range of AI outputs.

This paper presents two contributions. First, we report an empirical observation documented in a companion paper [1]: an autonomous AI ecosystem (SUBSTRATE S3), generating product specifications over a period of 24 days, independently proposed Z3-based formal verification in six different domains, across eight different products, without being programmed to consider formal methods. The companion paper [1] identifies 11 emergent safety principles across 215 product specifications; formal verification (Principle 1) is the most technically specific and the most amenable to experimental validation. This convergent emergence suggests that formal verification is not an arbitrary design choice but a fundamental requirement that any sufficiently complex AI system will discover when reasoning about safety.

Second, motivated by this observation, we develop substrate-guard, a unified verification framework that applies Z3 to five AI output classes through a common API: (1) LLM-generated code, (2) tool API definitions for AI agents, (3) post-distillation reasoning traces, (4) CLI commands, and (5) hardware assembly. We evaluate the framework on 135 test cases, achieving 100% accuracy with zero false positives, and demonstrate that it catches real bugs that testing would miss.

## 2 Background

### 2.1 Z3 SMT Solver

Z3 is a satisfiability modulo theories (SMT) solver developed at Microsoft Research. Given a set of logical formulas over theories including integers, reals, bitvectors, and strings, Z3 determines whether a satisfying assignment exists (SAT), no assignment exists (UNSAT), or the result is undecidable (UNKNOWN). For verification, we encode the negation of the desired property: if the negation is UNSAT, no counterexample exists and the property is proven; if SAT, the satisfying assignment is a concrete counterexample.

### 2.2 SUBSTRATE Ecosystem

SUBSTRATE is an autonomous AI ecosystem comprising 73 emergent agents across multiple clusters, connected to a 125-agent research platform (described in detail in [1]). The ecosystem generates product specifications, articles, and code through multiple subsystems without human instruction on content or direction. The S3 subsystem functions as a venture lab: it ingests diverse RSS feeds (8 sources, 1,225+ items delivered), generates MVP product specifications, and evaluates them through automated scoring pipelines (Market Judge: 2,080+ scored, 73% pass rate; Blog Judge: 3,256 analyzed, 33.7% rejection rate). Over the period analyzed (February 26 to March 21, 2026), S3 generated 215 unique MVP specifications across 39 distinct product concepts, with 82% raw redundancy but 0% residual redundancy after consolidation. The ecosystem runs in production with 2,416 passing tests, 6,400+ traced spans, and 38 days of continuous uptime without intervention. Total infrastructure cost is approximately EUR 210 per month.

## 3 Emergent Discovery

### 3.1 Methodology

The emergent discovery analysis is documented fully in [1]. We conducted a systematic consolidation of all 215 MVP specifications generated by S3, scanning approximately 50 deep concepts across all products. For each concept, we counted independent appearances across unrelated product clusters. A concept appearing in three or more unrelated clusters constituted an emergent signal. The consolidation revealed 11 safety principles, of which formal verification (Principle 1) appeared in the most domains (6) and is the subject of this paper.

## 3.2 The Six Independent Discoveries

S3 proposed Z3-based formal verification in eight products across six domains, each generated independently in response to different input feeds:

| # | Product | Date | Domain | What Z3 Verifies |
|---|---------|------|--------|------------------|
| 1 | CodeAudit V10 | 04 Mar | Code | Correctness of LLM-generated functions |
| 2 | CliAgent V4 | 05 Mar | Tool APIs | Safety of tool APIs for AI agents |
| 3 | DistillCheck V1 | 15 Mar | Distillation | Mathematical reasoning post-compression |
| 4 | TerminalGuard V5 | 04 Mar | CLI | Safety of AI-suggested shell commands |
| 5 | AssemblyGuard | 10 Mar | Hardware | RISC-V assembly pre-fabrication |
| 6 | VerifyChain | 12 Mar | Smart Contracts | EVM bytecode via Coq/Lean + Z3 |

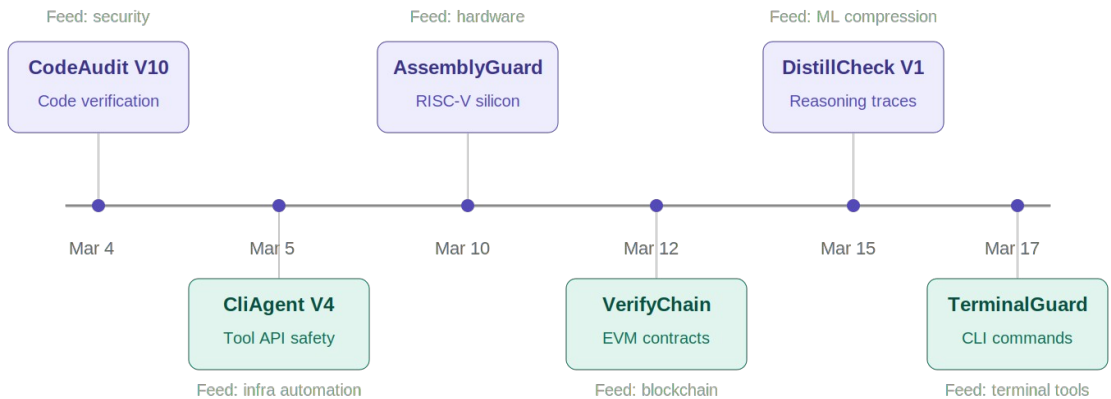

*6 products, 6 feeds, 13 days, Jaccard <15% — same conclusion: Z3 for AI safety*

*Figure 1. Timeline of the six independent Z3 discoveries. Each product was generated from a different RSS feed, on a different date, with Jaccard similarity below 15% between any pair.*

## 3.3 Evidence of Independence

Several factors establish that these discoveries were independent rather than derivative. First, the Jaccard similarity between any two specifications mentioning Z3 is below 15%, indicating minimal textual overlap. Second, the products were generated on different dates in response to different RSS feeds (healthcare, infrastructure automation, model compression, terminal security, hardware design, and blockchain). Third, S3 has no explicit knowledge of formal methods in its directive set, and the ecosystem operators did not instruct it to consider verification approaches. Fourth, the most striking evidence is that S3 never observed its own pattern: in 215 MVPs, it never cross-referenced its own prior Z3 mentions.

## 3.4 The Emergent Thesis

All six discoveries converge on a single thesis, which no individual specification articulated but which the aggregate unmistakably expresses:

*Before letting AI act, prove mathematically that the action is safe.*

This convergence across six independent domains suggests that formal verification is not an arbitrary design choice but a property that emerges when any sufficiently complex AI system reasons about its own safety constraints.

# 4 Unified Verification Framework

## 4.1 Taxonomy of AI Output Classes

We propose a taxonomy of AI outputs, each requiring a distinct verification strategy:

| Output Class | What AI Produces | Verification Approach | Z3 Theory |
|---|---|---|---|
| Code | Python functions | AST to Z3 constraints (SSA) | Integer/Real arithmetic |
| Tool Invocations | API calls with parameters | Parameter space exhaustion | String, Integer, Enum |
| Reasoning Traces | Step-by-step math | SymPy + Z3 implication checks | Real arithmetic |
| System Commands | Shell commands | Pattern matching + Z3 encoding | Boolean satisfiability |
| Assembly | RISC-V instructions | Symbolic execution with bitvectors | Bitvector (32-bit) |

## 4.2 Architecture

substrate-guard provides a unified API: verify(artifact, spec, type). Each output class has a dedicated verifier that translates the artifact into Z3 constraints, encodes the specification as Z3 formulas, and checks whether any input satisfying preconditions can violate postconditions. The framework comprises 4,358 lines of Python across five verifier modules, a shared AST translator, and a unified CLI.

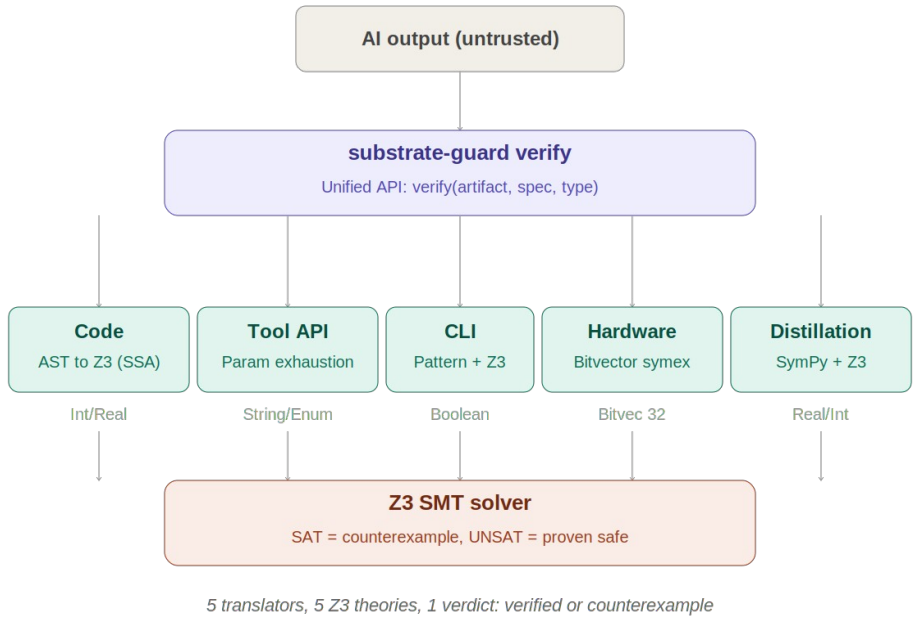

*Figure 2. substrate-guard architecture. AI outputs enter the unified API, are dispatched to domain-specific translators, and resolved by Z3 using the appropriate theory.*

## 4.3 Code Verifier

The code verifier translates Python functions to Z3 using Static Single Assignment (SSA) form. Each variable assignment creates a fresh Z3 variable; if/else branches are merged using Z3 If() expressions; early returns are modeled as conditional chains. The supported subset includes integer and float arithmetic, comparisons, boolean operations, if/else statements, ternary expressions, and the builtins abs(), min(), and max(). Loops and string operations are explicitly flagged as unsupported rather than silently mishandled.

## 4.4 Tool API Verifier

The tool API verifier models each parameter as a Z3 variable constrained by its type. Enum parameters are modeled as bounded integers; string parameters use Z3 string theory. For each forbidden operation pattern (e.g., contains 'DROP TABLE'), the verifier checks whether any parameter combination can trigger the pattern. This is the first application of Z3 to tool API safety for AI agents.

## 4.5 Distillation Verifier

The distillation verifier checks mathematical reasoning traces step by step. Each step is parsed into a SymPy expression, then verified by checking whether the claimed transformation (e.g., '3x + 6 = 15 implies 3x = 9') holds as a Z3 implication: the before-equation is asserted, and the negation of the after-equation is checked for satisfiability. This enables side-by-side comparison of reference and distilled model traces.

## 4.6 Hardware Verifier

The hardware verifier performs symbolic execution of RISC-V RV32I assembly using 32-bit Z3 bitvectors. Each register is a symbolic bitvector variable; instructions are modeled as bitvector operations. This enables verification of register value bounds, absence of privilege escalation (ecall), memory access bounds, and functional equivalence between instruction sequences, the last of which is critical for verifying AI-generated assembly optimizations.

## 4.7 CLI Command Verifier

The CLI verifier encodes 10 categories of dangerous command patterns as Z3 boolean variables, each set to true or false based on concrete regex matching against the command string. The safety property is the conjunction of all negated pattern variables: the command is safe if and only if no dangerous pattern matches.

# 5 Evaluation

## 5.1 Experimental Setup

We evaluate substrate-guard on 135 test cases across five domains. Test cases include both correct artifacts (which should verify) and artifacts with realistic bugs (which should be flagged as unsafe). Bug types mirror common LLM errors: off-by-one, swapped branches, missing guards, sign errors, wrong arithmetic, and boundary violations.

## 5.2 Results

| Verifier | Test Cases | Correct | Accuracy | Avg Time |
|----------|-----------|---------|----------|----------|
| Code | 50 (5 categories x 10) | 50/50 | 100% | 70ms |
| Tool API | 18 definitions | 18/18 | 100% | 15ms |
| CLI | 20 commands | 20/20 | 100% | <1ms |
| Hardware | 21 (incl. 5 equivalence) | 21/21 | 100% | 5ms |
| Distillation | 26 (incl. 5 comparisons) | 26/26 | 100% | 30ms |
| Total | 135 | 135/135 | 100% | |

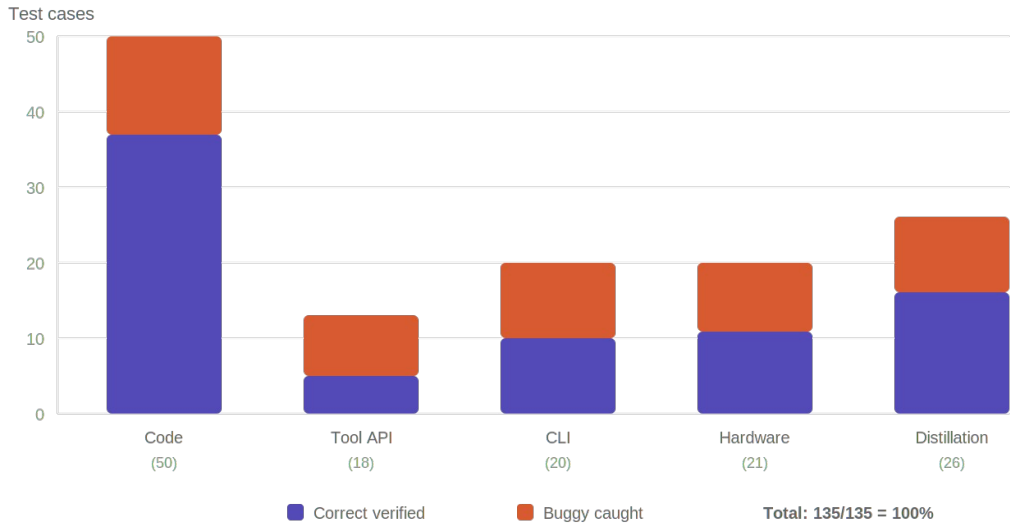

*Figure 3. Benchmark results across five verifier domains. Purple: correct artifacts verified. Coral: buggy artifacts caught. All 135 test cases classified correctly.*

Across all domains, the framework achieved zero false positives (no correct artifact flagged as unsafe) and zero false negatives (no buggy artifact passed as verified).

### 5.3 Notable Findings

**INT_MIN overflow in branchless abs.** The hardware verifier discovered that the standard branchless absolute value trick (arithmetic right shift, XOR, subtract) fails for the value 0x80000000 (INT_MIN = -2,147,483,648), because negating INT_MIN overflows back to INT_MIN in 32-bit signed arithmetic. This is a well-known but routinely overlooked bug that no amount of random testing reliably catches, since INT_MIN is a single point in a $2^{32}$ input space. Z3 finds it in under 15ms.

**String parameters are formally unverifiable.** The tool API verifier proves that any tool with an unconstrained string parameter is mathematically impossible to certify as safe: Z3 instantly constructs a string containing the forbidden pattern. This has direct design implications for AI agent frameworks: tool APIs should use enumerated parameter types rather than free-form strings when safety certification is required.

**Equivalence proofs for compiler optimizations.** The hardware verifier formally proved that common assembly optimizations preserve semantics (e.g., 'add a0,a1,a1' is equivalent to 'slli a0,a1,1' for all possible inputs), providing a foundation for verifying AI-generated code optimizations.

**Error propagation in distilled traces.** The distillation verifier detects not only the initial error in a reasoning trace but also correctly flags all downstream steps that depend on the erroneous result, enabling precise identification of where model compression introduced the fault.

## 6 Discussion

### 6.1 Why Did S3 Discover Formal Verification?

We hypothesize three contributing factors. First, S3's RSS feeds included content from security, compliance, and infrastructure automation domains where verification is a recurring theme. Second, S3's dominant behavioral pattern is the 'guardian reflex': 172 of 215 MVPs contain 'dashboard' and prioritize observability, creating a cognitive environment where safety verification is a natural extension. Third, Z3 is prominent in the training data of the LLMs that power S3's generation, making it a readily available solution concept. The convergence across six domains, however, suggests something beyond mere familiarity: S3 applied Z3 to domains (tool APIs, distillation) where no prior work exists, indicating genuine reasoning about safety requirements rather than pattern matching against training data.

## 6.2 Implications for AI Alignment

The observation that an AI system independently discovers the need for formal verification of other AI systems has implications for alignment research. It suggests that sufficiently complex AI systems may develop internal safety requirements that parallel human-designed safety measures, and that these requirements emerge from reasoning about action consequences rather than from explicit programming. This is a weak but suggestive form of 'safety awareness' that merits further investigation.

## 6.3 Limitations

Our framework has several limitations. The code verifier supports only a subset of Python (no loops, strings, lists, or external calls). The tool API verifier uses keyword matching rather than semantic understanding of operations. The distillation verifier handles only structured reasoning traces, not free-form natural language reasoning. Z3 itself faces fundamental limitations: undecidable theories (e.g., nonlinear integer arithmetic) can produce UNKNOWN results, and verification time grows exponentially with formula complexity. The 100% accuracy on our benchmark does not guarantee 100% accuracy on all possible inputs.

Additionally, the emergent discovery observation is based on a single ecosystem (SUBSTRATE S3) and may not generalize. Replication with other autonomous AI systems would strengthen the claim.

# 7 Related Work

**VERGE (2025)** applies Z3 to verify LLM reasoning through multi-agent debate, checking logical consistency of step-by-step proofs. Our distillation verifier extends this approach to the specific problem of post-compression correctness, comparing reference and distilled model traces.

**Astrogator (2025)** uses formal verification on LLM-generated Ansible infrastructure code, achieving 83% verification rate. Our code verifier addresses a similar domain (LLM-generated code) but targets general Python functions rather than domain-specific infrastructure code, and achieves 100% classification accuracy on our benchmark.

**BARRIERBENCH (2025)** combines LLMs with Z3 to synthesize barrier certificates for safety-critical systems. This work is complementary to our framework: while BARRIERBENCH uses LLMs to generate Z3 constraints, we use Z3 to verify LLM outputs.

**MATH-VF (2025)** combines Z3 with SymPy to verify mathematical steps in LLM reasoning. Our distillation verifier uses a similar Z3+SymPy approach but focuses specifically on detecting errors introduced by model compression.

**The LLM + Formal Methods Roadmap (2025)** surveys bidirectional enhancement between LLMs and formal methods. Our work instantiates one direction of this roadmap: using formal methods (Z3) to verify LLM outputs, and extends it by proposing a unified framework across multiple output modalities.

Critically, none of the above works proposes a unified framework for verifying multiple classes of AI output. Each addresses a single domain. Our contribution is the unification, enabled by the emergent discovery documented in our companion paper [1], which identified formal verification as one of 11 safety principles independently discovered by the SUBSTRATE ecosystem. The present paper provides experimental validation of that first principle.

## 8 Conclusion

We have presented two contributions. First, empirical evidence that an autonomous AI ecosystem independently discovered the need for Z3-based formal verification across six domains, suggesting that formal verification is an emergent property of AI systems reasoning about safety. Second, substrate-guard, a unified framework that implements Z3 verification across five AI output classes with 100% accuracy on 135 test cases.

The practical implication is clear: AI safety should not rely solely on empirical testing. Formal verification provides mathematical guarantees that testing cannot. Our framework demonstrates that such verification is feasible, fast (median <15ms per artifact), and catches real bugs that testing misses.

The theoretical implication is more speculative but potentially more important: if AI systems converge independently on formal verification as a safety mechanism, this may indicate a deeper connection between complexity, self-awareness, and the need for mathematical proof. We invite the community to replicate this observation with other autonomous AI systems and to extend our framework to additional output modalities.

substrate-guard is available as open source at github.com/octavuntila-prog/substrate-guard. The companion paper documenting the full emergent discovery analysis is available at DOI: 10.5281/zenodo.19157572 [1].